\documentclass[
    ,final            
  ]
  {aipproc}

\layoutstyle{8x11double}
\renewcommand\XFMtitleblock{%
 \XFMtitle
 \let\XFMoldpar\par
 \def\par{\XFMoldpar\def\par{\space
            for the VERITAS Collaboration \XFMoldpar}}%
  \XFMauthors
  \let\par\XFMoldpar
  \begin{center}
  \XFMaddresses
\end{center}
  \XFMabstract
  \vspace{5pt}%
  \XFMkeywords
  \XFMclassification
 }

\begin{document}

\title{VERITAS Distant Laser Calibration and Atmospheric Monitoring}


\author{C. M. Hui}{
  address={University of Utah, Dept. of Physics, 115 S 1400 E, Salt Lake City,
  UT 84112, USA}, 
  email={cmhui@physics.utah.edu}
}

\begin{abstract}
As a calibrated laser pulse propagates through the atmosphere, the intensity
of the Rayleigh scattered light arriving at the VERITAS telescopes can be
calculated precisely.  This allows for absolute calibration of imaging
atmospheric Cherenkov telescopes (IACT) to be simple and straightforward.  In
these proceedings, we present the comparison between laser data and simulation
to estimate the light collection efficiencies of the VERITAS telescopes, and
the analysis of multiple laser data sets taken in different months for
atmospheric monitoring purpose. 
\end{abstract}

\maketitle


\section{Introduction}
VERITAS is an array of IACTs located at the Fred Lawrence Whipple Observatory
on Mount Hopkins in Arizona \cite{Holder08}. It consists of four 12m
reflectors, each with a camera comprising 499 photomultiplier tubes arranged
in a hexagonal lattice covering a field of view of $3.5^\circ$.  Calibration
of the VERITAS telescopes is done via measurements of individual elements as
described in \citet{Hanna07}.  A more direct approach to measure the effective
light collection area of the VERITAS telescopes is first proposed by
\citet{Shepherd05} using scattered light from a calibrated laser pulse that is
detectable by the telescopes and can be simulated easily.  As the laser shot
travels upward through the atmosphere, the laser light undergoes Rayleigh
scattering from dipoles, and Mie scattering from particles similarly sized as
the laser wavelength or bigger \cite{Kyle}.  If the telescopes are
pointing at high elevation (the highest line in figure \ref{rl}), Rayleigh
scattering would dominate after attenuation from the aerosol layer.  For
mid-range elevation, the scattered laser light would have gone through a
longer distance in the aerosol layer and attenuated more than at the higher
elevation.  If the telescopes are pointed at low elevation and intercepting
scattered light coming from within the aerosol layer (the lowest line in
figure \ref{rl}), Mie scattering becomes important. 

Rayleigh scattering can be simulated accurately while Mie scattering depends
on the aerosol conditions and properties at the particular time.  Therefore
multiple measurements over a range of altitudes throughout the observing
season are necessary to monitor atmospheric changes and to estimate the
effective light collection area of the telescopes. 

\section{Laser Setup $\&$ Data Acquistion}

\begin{figure}[t]
  \includegraphics[width=0.45\textwidth]{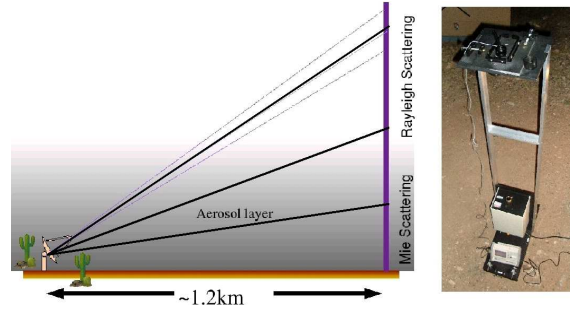}
  \caption{Cartoon demonstration of laser setup and a picture of the laser.}
\label{rl}
\end{figure}

A 300\,$\mu J$ nitrogen laser with 4\,ns pulse width and 337nm wavelength is
mounted on a movable rack with flexible beam collimation and intensity
adjustment (see figure \ref{rl}).  The laser is fired pointing at zenith $\sim
1.2$ km away from the VERITAS telescope array.  The array is aimed at a range
of elevation from $20^\circ$ to $60^\circ$ (altitudes 0.5 km to 2.2 km).  Both
the laser and the array are externally triggered by GPS pulsers such that each
recorded event contains an image of the laser shot. The flash-ADC (FADC)
recording window is lengthened to 244 samples (488 nanoseconds) to maximize
the recording time of the scattered laser light moving across the camera from
3.5 to $12.5^\circ \mu s^{-1}$ depending on the elevation.  

\begin{figure}[t]
  \includegraphics[width=0.95\textwidth]{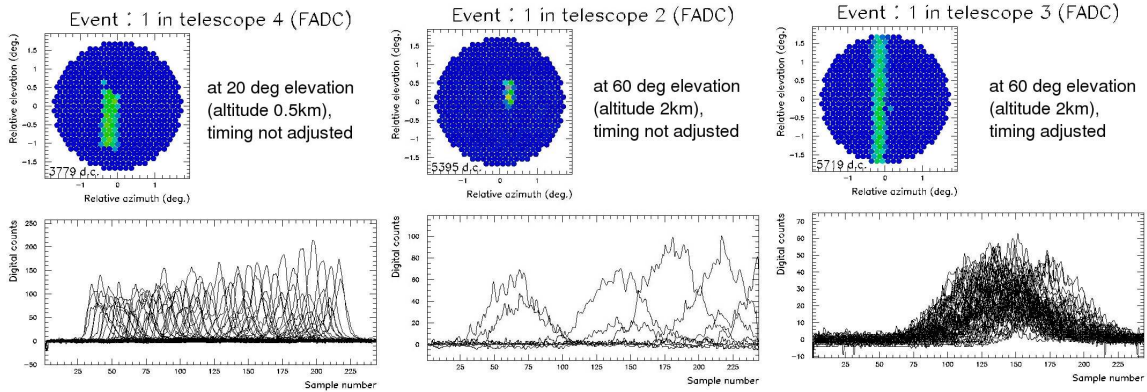}
  \caption{FADC traces of a laser event at different elevation.  At higher
    elevation, the number of pixels triggered is reduced due to geometrical
    effects.  With timing adjustments, the scattered laser light can be
    captured with the entire length of the camera.}
\label{rlcam}
\end{figure}

Due to the limited recording time in data acquisition and geometric effects,
the length of the laser signal trace increases with altitude (see figure
\ref{rlcam}).  To alleviate this, the pixels are configured to read at
different memory depths of the recorded trace for altitude greater than
1.5\,km such that all recorded events from any altitude have more than 20
pixels in the image.

\section{Analysis $\&$ Simulation}
The laser analysis algorithm is modified from the regular data analysis
used in VERITAS described in \citep{Acciari08}.  The FADC trace is
convolved with a fixed 125-sample wide window and the location of the
integration window is determined by the peak of the integration.  If the
integration window falls outside of the FADC recording window, the trace is
considered truncated and discarded from the analysis.  For the remaining
traces that do not resemble a background signal and have an integrated digital
counts (dc) more than 180, those are added to the image sum.  The
Linear Image Brightness Density (LIBD) is the image sum divided by the image
length.  

Rayleigh scattering simulation with GrISU(tah) \cite{grisu} simulation package
provides the expected VERITAS detector output from the Rayleigh-scattered
light.  The detector output is then analyzed by the algorithm described above.
At higher altitude where Mie scattering is less important, the LIBD from the
simulation is matched to data by adjusting the light collection efficiency
constants in the telescopes model.  Combined with tracing of the
Rayleigh-scattered light in simulation, we can derive the effective light
collection area of the telescopes. 

\section{Data $\&$ Results}

\begin{figure}
  \includegraphics[width=.4\textwidth]{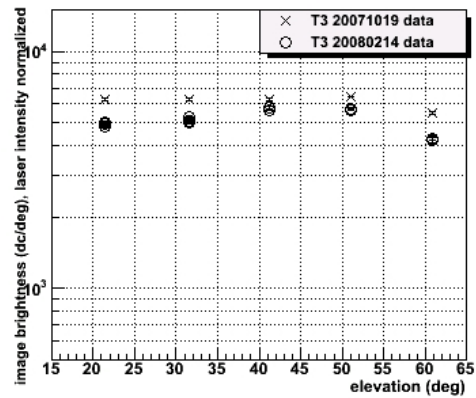}
  \caption{Linear image brightness density measured in two different nights,
    laser intensity has been normalized.  Both nights are comparable to each
    other, suggesting similar atmospheric conditions and minimal changes in
    the telescopes on both nights.}
\label{rldata}
\end{figure}

Two sets of data were taken in October 2007 and Feburary 2008.  To account for
the difference in laser intensity, it is divided from the LIBD.  Both nights
has comparable LIBD with laser intensity normalized, suggesting little changes
in the local atmosphere and the telescopes between the two nights (see figure
\ref{rldata}).  More frequent distant laser runs are being planned for regular
monitoring of the local atmosphere and of the array. 

\begin{figure}
  \includegraphics[width=.4\textwidth]{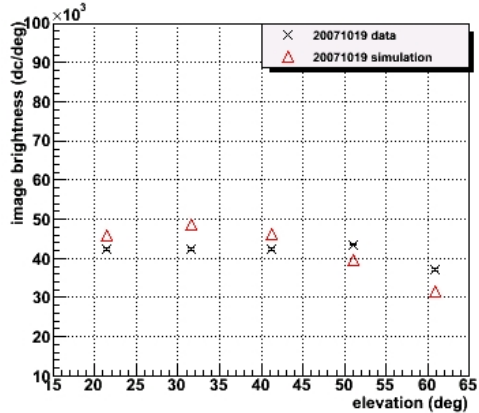}
  \caption{Linear image brightness density comparison between data and
    simulation.  Data and simulations are well-matched above $50^\circ$
    elevation.} 
\label{rlsim}
\end{figure}

\begin{figure}
  \includegraphics[width=.4\textwidth]{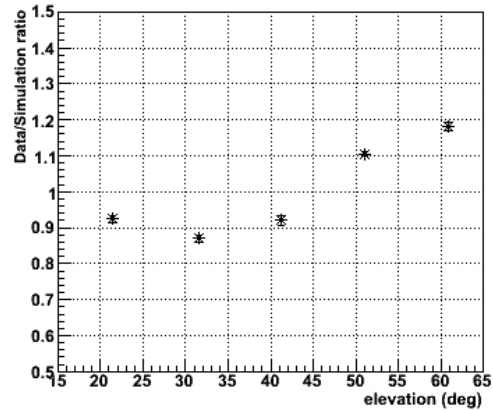}
  \caption{Linear image brightness density ratio between data and simulation.  At low
    elevations the ratio is lower due to additional Mie scattering not
    accounted for by the simulation.}
\label{rleff}
\end{figure}

Outside the aerosol layer, Rayleigh scattering dominates over Mie scattering.
In figure \ref{rlsim}, the LIBD of our measurement and simulation are shown
together for comparison.  At high elevation, the data is brighter than the
simulation; while at low elevation, the scattered light travels a longer
distance inside the aerosol layer and undergoes more Mie scattering, lowering
the light intensity reaching the telescopes.  Figure \ref{rleff} shows the
ratio of LIBD between data and simulation as a function of elevation.  This
ratio can be affected by the accuracy of the telescopes model used in our
simulation and by the amount of Mie scattering that is not accounted for in
our simulation.  The high elevation points are affected least by Mie
scattering unaccounted for in simulation, allowing us to adjust our telescopes
model in simulation accordingly. 

By comparing data at high elevation and the calculated density of Rayleigh
scattered light reaching the telescopes from simulation, we can estimate the
effective light collection area of individual telescopes.  The recorded LIBD at
$60^\circ$ elevation for one telescope is $3.7\times10^4$\,dc\,$deg^{-1}$ per
event, and there were 515\,photons\,$m^{-2}deg^{-1}$ per simulated event at the
same elevation.  Therefore the ratio of how many digital counts per photon for
the telescope's light collecting area is 72\,dc\,$m^2 photon^{-1}$.  Single
photoelectron measurement derived from the nightly flatfielding laser runs
\cite{Hanna07} is $\sim 5$\,dc/photon.  Combined with the light collecting
ratio, the effective light collection area of a telescope is $\sim
14$\,$m^2$.  

The effective light collection area can also be approximated by using
indiviual elements' measurements.  The VERITAS telescopes each have a mirror
area of $\sim 110$\,$m^2$, quantum efficiency at laser wavelength is measured
to be 0.18, mirror reflectivies at laser wavelength is 0.92, and camera
efficiency is 0.81.  These measurements yield an effective collection area of
15\,$m^2$, which is comparable to the area derived from the distant laser
measurement. 

\section{Summary}
A calibrated laser pulse fired at zenith at some distance away from IACTs can
be used for absolute calibration and for monitoring of both the telescopes'
conditions and the atmosphere at the same time.  We presented 2 nights of data
that shown similar linear image brightness density, suggesting few changes in
the atmosphere and the conditions of the telescopes during that time.  With
more measurements in the future, we hope to characterize the effect of
aerosols on light collection of IACTs.  Preliminary estimation of light
collection efficiency of one of the telescopes shows our expectation of
telescope performance matches with what was recorded, and the effective light
collection area is measured to be $\sim 14$\,$m^2$. 


\begin{theacknowledgments}
This research is supported by grants from the U.S. Department of Energy, the
U.S. National Science Foundation, and the Smithsonian Institution, by NSERC in
Canada, by PPARC in the UK and by Science Foundation Ireland. 
\end{theacknowledgments}

\end{document}